\documentclass[twocolumn]{revtex4-1}
\usepackage[english]{babel} 
\usepackage{amsmath}
\usepackage{amssymb}
\usepackage{graphicx}
\usepackage{epstopdf}
     
\begin{document}
	
\title{Effect of edge vacancies on localized states in semi-infinite zigzag graphene sheet}

\author{Glebov A.A.$^{a,b}$, Katkov V.L.$^{b}$, and Osipov V.A.$^{b}$}
\affiliation{$^{a}$Moscow Institute of Physics and Technology, 141700, Dolgoprudny, Moscow region, Russia}
\affiliation{e-mail: glebov.atth@gmail.com}
\affiliation{$^{b}$Bogoliubov Laboratory of Theoretical Physics, Joint Institute for Nuclear Research, 141980 Dubna, Moscow region, Russia}

\begin{abstract}
The effect of vacancies on the robustness of zero-energy edge electronic states in zigzag-type graphene layer  is studied at different concentrations and distributions of defects. All calculations are  performed by using the Green's function method and the tight-binding approximation. It is found that the arrangement of defects plays a crucial role in the destruction of the edge states. We have specified a critical distance between edge vacancies when their mutual influence becomes significant and affects markedly the density of electronic states at graphene edge.	

\end{abstract}

\maketitle

\begin{center}
	\textbf{Introduction}
\end{center}	

Localized zero-energy edge electronic states were theoretically predicted in graphene ribbons and semi-infinite sheets with a crystallographically clean zigzag-type termination~\cite{klein,au1,jpn}. 
These states are characterized by a high electronic density (DOS) at the Fermi level, which was experimentally observed using scanning tunneling microscopy \cite{au2,au3}.

Generally, vacancies appear during the synthesis of graphene \cite{au7} and in reality one has to deal with nonideal structures. Vacancies that are located on the edge atoms of graphene sheet disturb the lattice structure and can affect the stability of edge states. In turn, this will drastically influence the main characteristics of nanoelectronic devices. The known example is an attempt to use the effect of spin polarization in graphene nanoribbons for spintronics applications~\cite{au4,au5,au6}. The spin polarization originates from the edge states that introduce a high density of state at the Fermi energy and is found to be greatly suppressed in the presence of edge defects and impurities. It was shown that the spin suppression is caused by the reduction of DOS at the Fermi energy and the GNR becomes nonmagnetic at a critical concentration of one edge defect per 1 nm~\cite{au9}. This makes practical applications in spintronic devices rather challenging.

Edge states are expected to play an important role in graphene nanoelectronic applications providing a specific tunneling current in graphene-based tunnel junctions \cite{au10,au13,au17}. Evidently, vacancies will influence the main characteristics of nanoelectronic devices operating on the basis of tunnel current through these states. The aim of this work is to study the influence of both the concentration and different location (normal, uniform, periodic) of vacancy defects on the stability of edge states in zigzag-type semi-infinite graphene sheet.\\

\begin{center}
	\textbf{The model}
\end{center}

\begin{figure}[t!]
	\begin{center}
		\includegraphics[width=8cm]{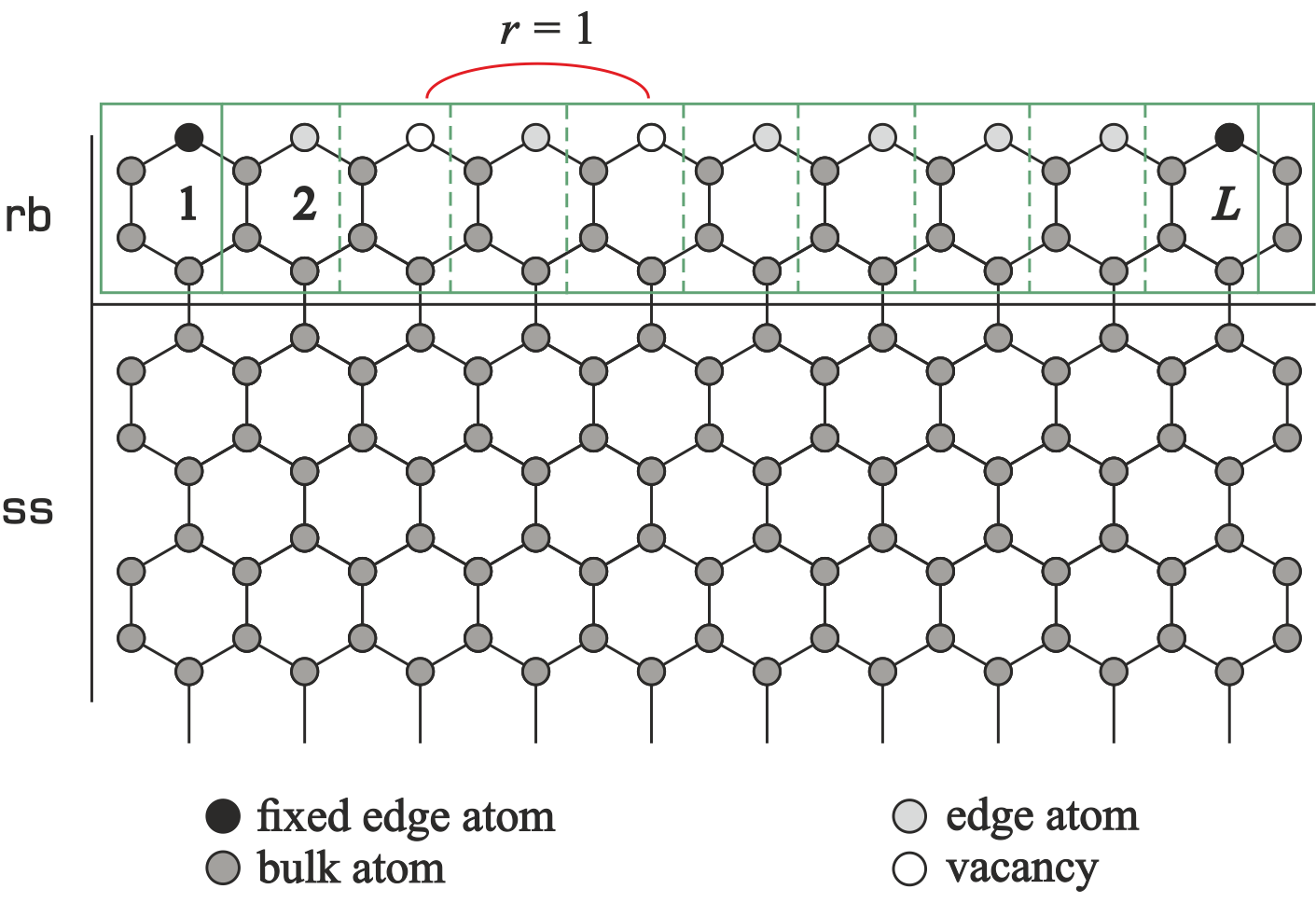}
	\end{center}
	\caption{Graphene lattice is divided into two blocks: $ss$ - semi-infinite graphene layer, $rb$ - graphene ribbon. The pictured cells are numbered from 1 to $L$, $r$ defines the number of atoms between vacancies. The density of states is calculated for edge atoms only.}
	\label{pon0}
\end{figure}

The calculations are performed by using the Green's function method and the tight-binding approximation. 
Graphene sheet can be split into two parts: a semi-infinite list ($ss$) and a ribbon ($rb$) containing the vacancies (see Fig.~\ref{pon0}). 
The ribbon is divided into $L$ parts with $L$ taken the values 12, 49, 100, and 400. 
Vacancies may appear on any positions of edge atoms except 1 and $L$. The B3LYP-based analysis within the ORCA~\cite{orca} shows that while short graphene ribbons containing hydrogen at the edges tend to bending, the wider ribbons are much more stable. Therefore, we consider semi-infinite graphene sheet as a planar one.
This structure is described by the Hamiltonian consisting of two parts: $H_{ss}$ and $H_{rb}$.
Vacancy defects are taken into account by replacement of zero value in the diagonal cell of the matrix $H_{rb}$ for the corresponding edge atom to infinity~\cite{au16}. 
The density of states is written as  

\begin{equation}
DOS(\epsilon)=(-1/\pi)\rm{Im}\Sigma(\textit{G}_{\textit{n,n}})/\textit{L},
\end{equation}
where $\epsilon$ is the energy and ${G}_{n,n}$ is an element of the retarded Green's function for the ribbon given by

\begin{equation}
G_{rb}(\epsilon)=[(\epsilon+i0^{+})\textbf{1}-H_{rb}-Tg_{ss}T^{\dagger}]^{-1},
\end{equation}

\begin{equation}
g_{ss}(\epsilon)=[(\epsilon+i0^{+})\textbf{1}-H_{ss}]^{-1}.
\end{equation}
Here $T$ is  the interaction matrix between the $ss$ and $rb$, $g_{ss}$ is the retarded Green's function for the sheet, $\textbf{1}$ is the identity matrix. $g_{ss}$ is calculated by using of the iterative algorithm described in Refs. \cite{au14, au15}. The energy is expressed in units of the interaction parameter $t$. Figs. 2-7 show the local (LDOS) and total (TDOS) density of states for edge atoms in the case of $L=100$. Notice that similar results were obtained for all considered $L$. 

\begin{center}
	\textbf{Characteristic range of mutual influence}
\end{center}

\begin{figure}[t!]	
	\begin{center}	
		\includegraphics[width=8cm]{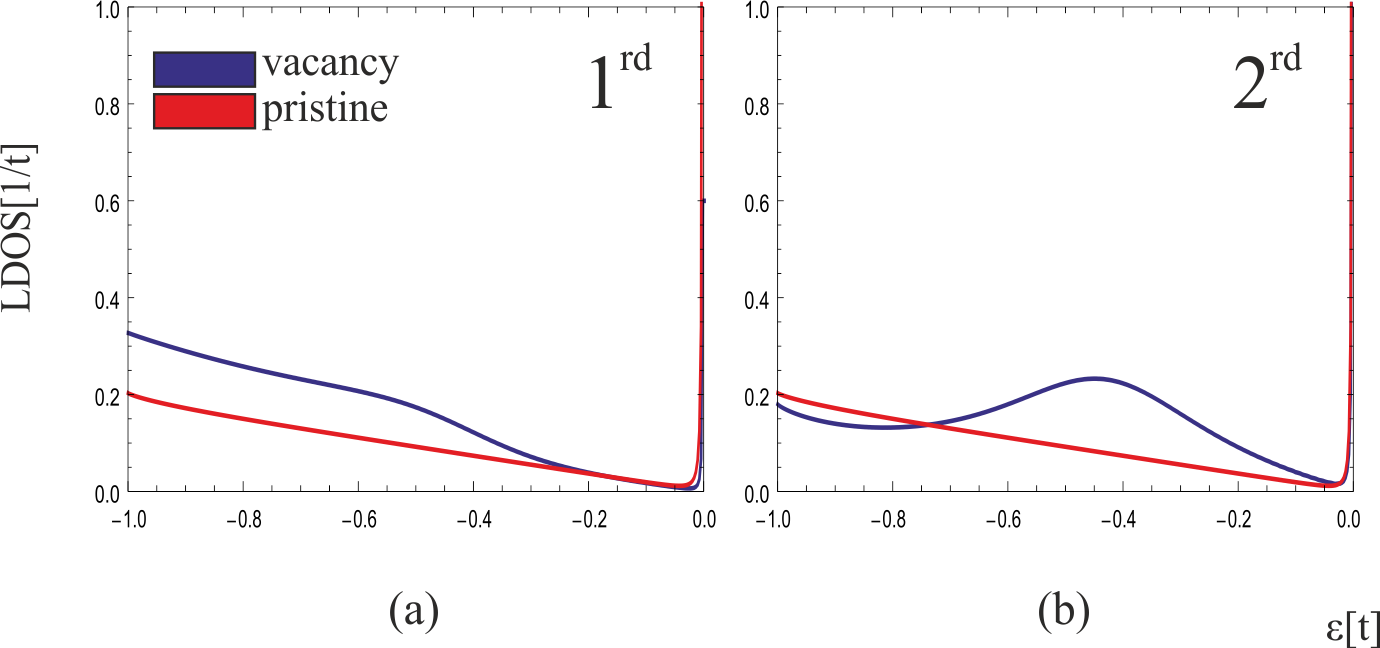}
		\includegraphics[width=8cm]{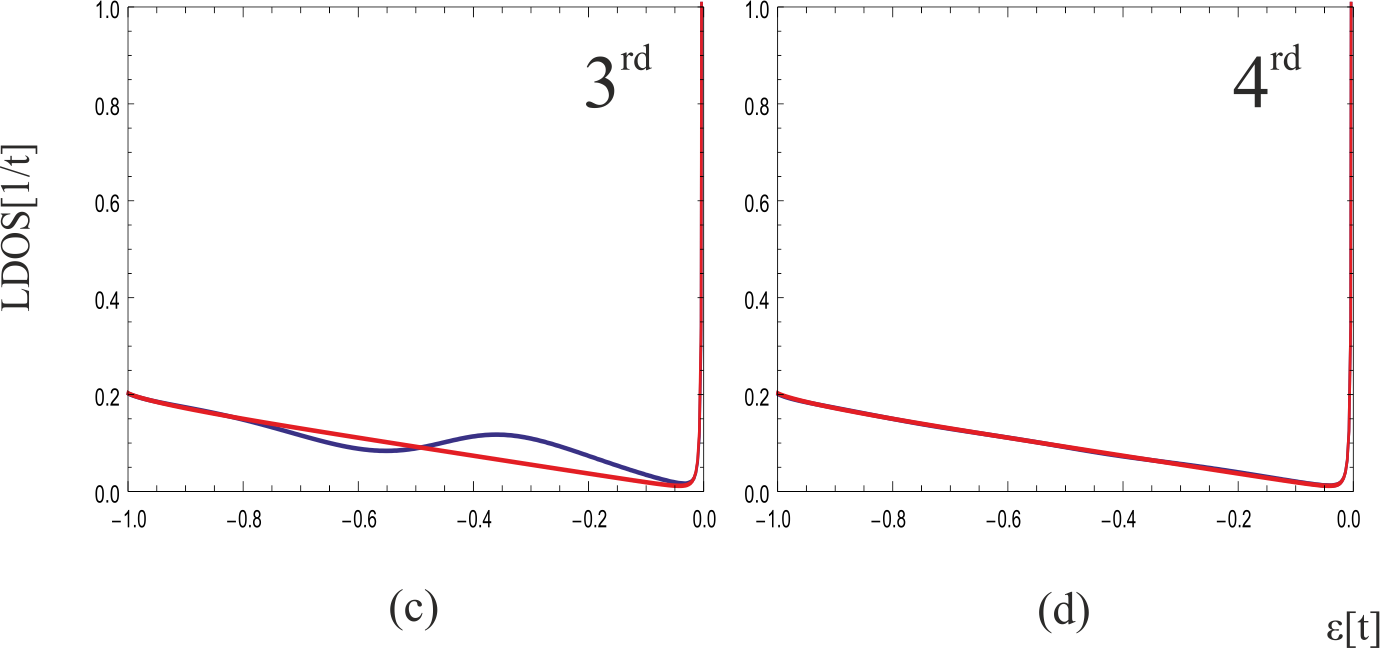}	
	\end{center}
	\caption{LDOS for four nearest to a single vacancy atoms at zigzag edge. Defect-free case is marked in red.}	
	\label{pon}
\end{figure}

\begin{figure}[t!]
	\begin{center}	
		\includegraphics[width=8cm]{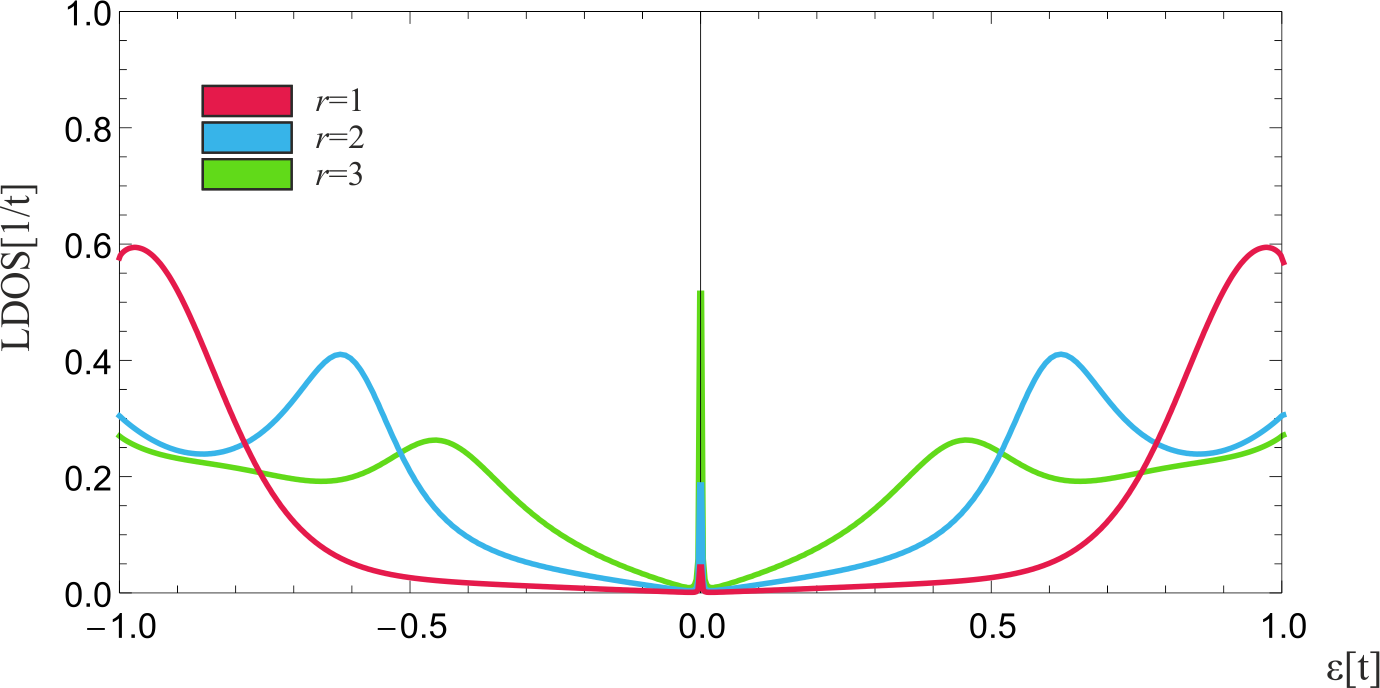}	
	\end{center}
	\caption{LDOS of edge atoms located between two vacancies for different $r$. Subpeaks are associated with the presence of defects.}	
	\label{pon1}
\end{figure}

For pristine semi-infinite graphene with zigzag termination, the density of electronic states is found to have a characteristic peak. Our study shows that a single vacancy does not affect the TDOS. This finding agrees with the results of Ref.~\cite{au19} where it was shown that a single vacancy has no effect on the localized zero-energy states of zigzag-terminated edge ribbons. At the same time, LDOS is found to be sensitive to a single vacancy at a distance up to three nearest neighbors (3NN) (see Fig.~\ref{pon}). 
This can be considered as a characteristic range where a single vacancy has an impact. 
Let us add the second defect and denote by $r$ the number of atoms between vacancies.
The defects are located inside the above defined characteristic range.  
As is seen in Fig. \ref{pon1}, LDOS turns out to be redistributed in the area between defects. 
The height of the main peak associated with the edge state in pristine graphene reduces an additional subpeak emerges which can be explained by mutual influence of defects.

\begin{figure}[t!]
	\begin{center}	
		\includegraphics[width=8cm]{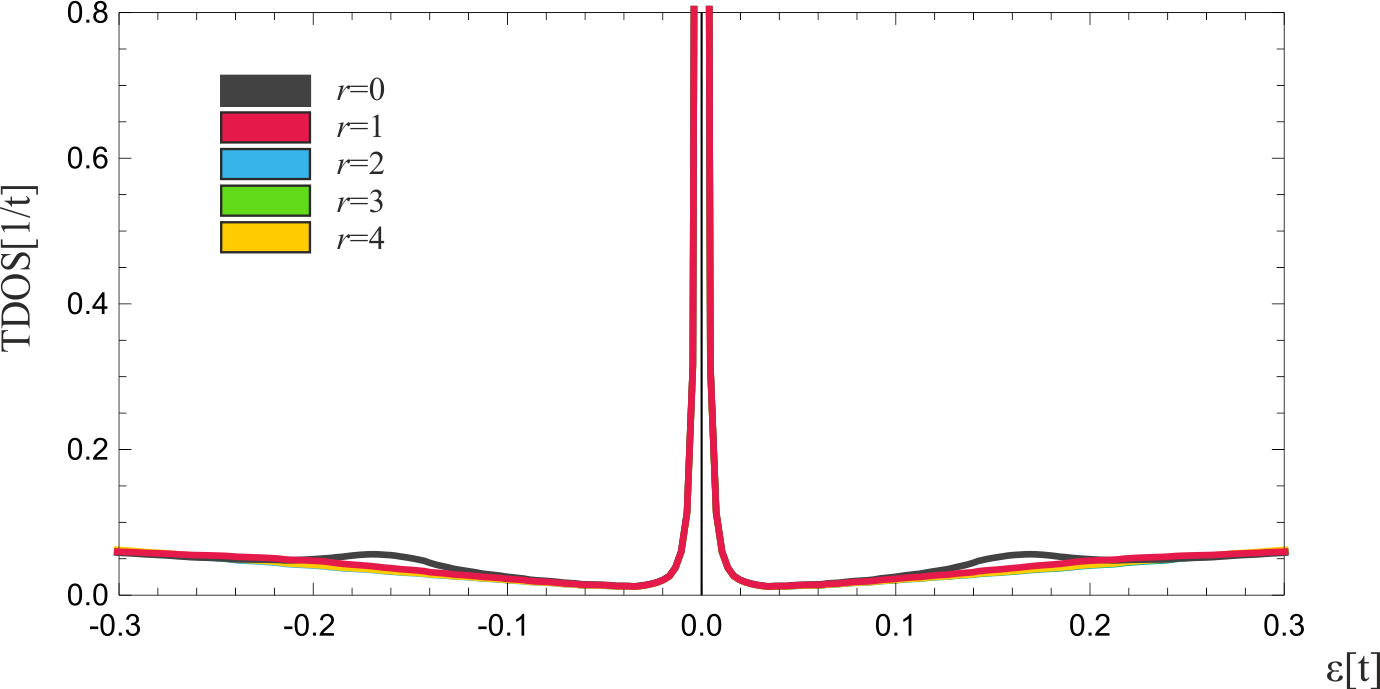}	
	\end{center}
	\caption{TDOS of edge atoms for different $r$. Subpeaks are associated with the divacancy ($r$=0). Vacancy pair with $r\geq1$ does not affect the TDOS.}	
	\label{pon3}
\end{figure}

A divacancy ($r=0$) changes the LDOS as in the case of single vacancy, but subpeaks associated with defects are revealed in the TDOS (see Fig. \ref{pon3}, black line). 
Our analysis shows that subpeaks do not appear for other values of $r$. What is important to note, vacancy pair does not affect the stability of edge states.

\begin{center}
	\textbf{Effect of concentration and distribution pattern of edge vacancies}
\end{center}

We will consider the cases of different (30, 50 and 70\%) vacancy concentrations. 

(i) Normal distribution. We use the normal distribution~\cite{au20} to simulate a situation when an array of vacancies is mainly situated in the middle of the edge. Defects are positioned by making use of a random number generator. 
Each atomic position is assigned a weighting factor in accordance with the normal distribution
\begin{equation}
	f(x)=\dfrac{1}{\sigma\sqrt{2\pi}}e^{-\dfrac{(x-\tilde{x})^{2}}{2\sigma^{2}}},
\end{equation}
where $x$ denotes the number of the cell, $\tilde{x}=L/2$ is the central cell which corresponds to the middle of the graphene ribbon, and $\sigma$ is the dispersion of the distribution.
We consider two different cases with $\sigma_{1}=L/4$ and $\sigma_{2}=3L/4$. 

\begin{figure}[t!]	
	\begin{center}	
		\includegraphics[width=8cm]{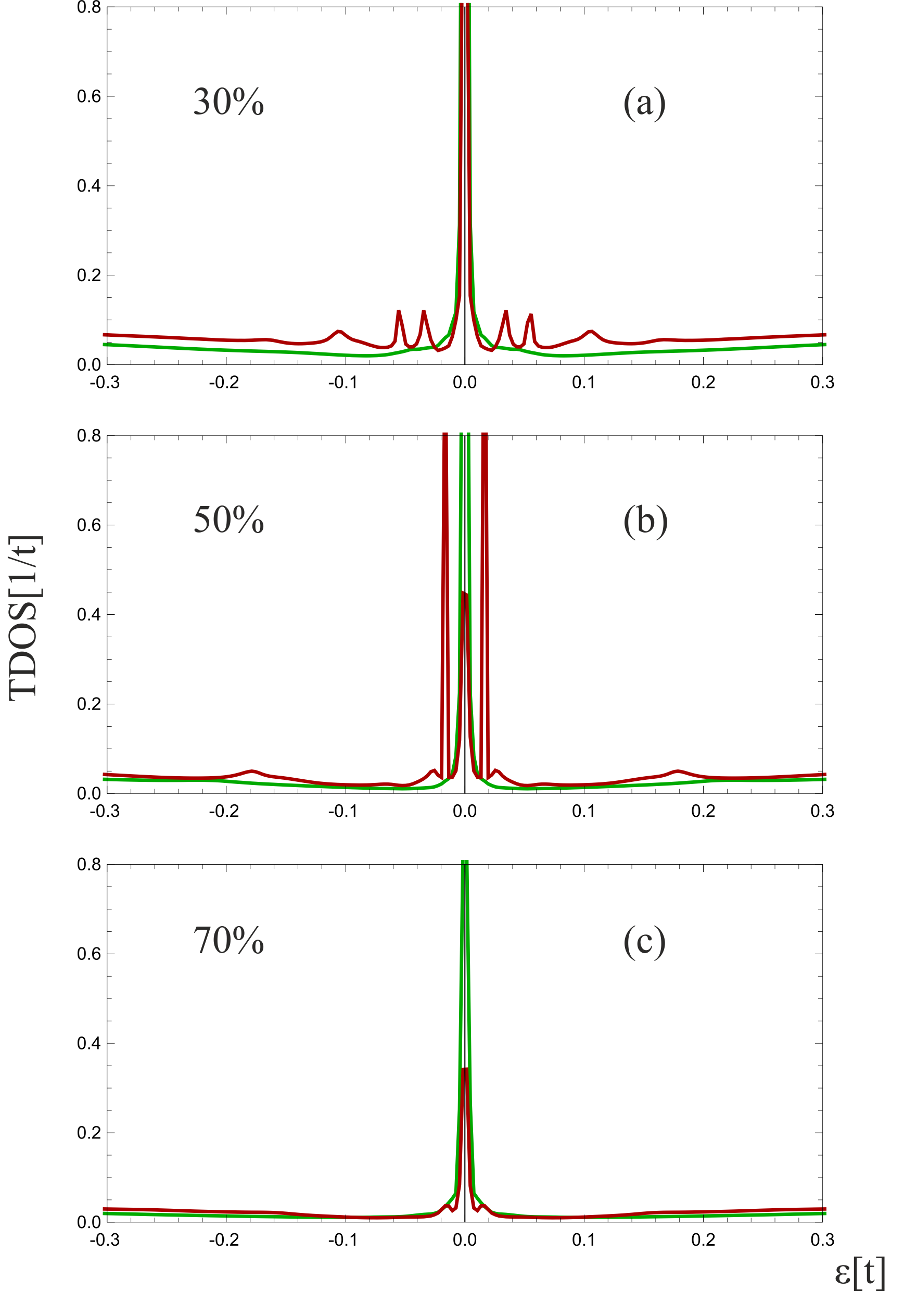}  	
	\end{center}	
	\caption{TDOS at different vacancy concentrations for a normal distribution with the dispersion parameters $\sigma_{1}$ (green line) and $\sigma_{2}$ (red line). The peak at the Fermi level decreases significantly at high dispersion. The edge state survives in all cases.}	
	\label{pon4}
\end{figure}

Increasing $\sigma$ enlarges the number of options for the distribution of defects relative to the middle of the graphene ribbon. Fig.~\ref{pon4} shows the TDOS with different concentrations and values of $\sigma$. 
\begin{figure}[t!]	
	\begin{center}	
		\includegraphics[width=8cm]{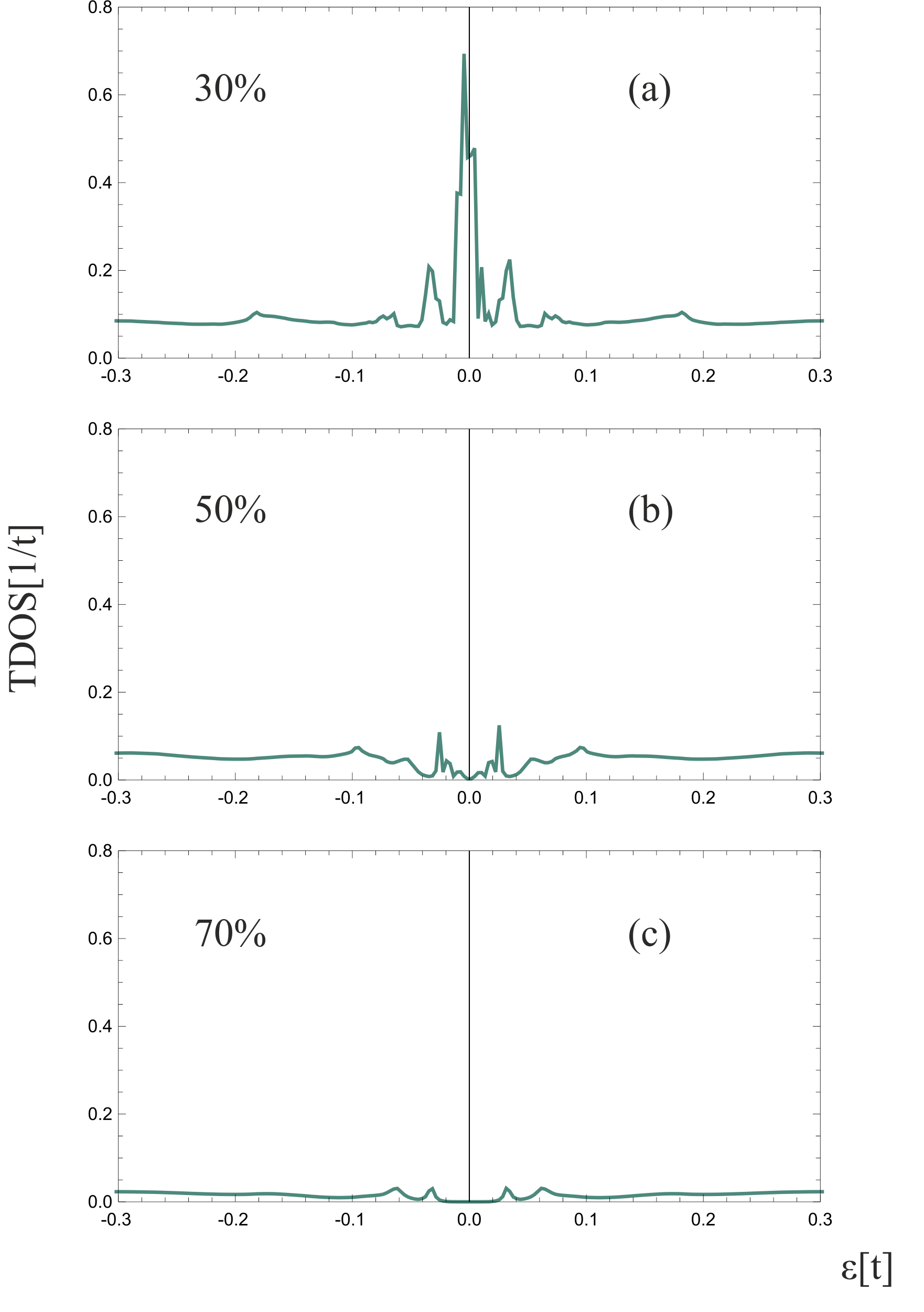}  
	\end{center}	
	\caption{TDOS at different vacancy concentrations for a uniform distribution. Averaged values are shown for ten random allocations of vacancies. The edge state disappears at 50\% vacancy concentration and higher.}	
	\label{pon6}
\end{figure}
For  $\sigma_{1}$, the reduction of the peak near the Fermi level takes place and an additional subpeak emerges while the edge state remains unbroken.
In the case of  $\sigma_{2}$, the edge states are markedly reduced at 50 and 70\% concentration.
Notice that in both cases the edge state disappears with further increase of the vacancy concentration.

(ii) Uniform distribution of vacancies. 
For this type of distribution, the edge state is found to destroy at 50\% vacancy concentration (see Fig. \ref{pon6} (b)). 
As an explanation, there is a large number of free-standing vacancies, which are located between one or two atoms. Its peculiarity lies in the fact that the height of subpeaks is greater than the peak at the Fermi level. Further increase of vacancy concentration leads to disappearance of the edge state.

\begin{figure}[t!]	
	\begin{center}
		\includegraphics[width=8cm]{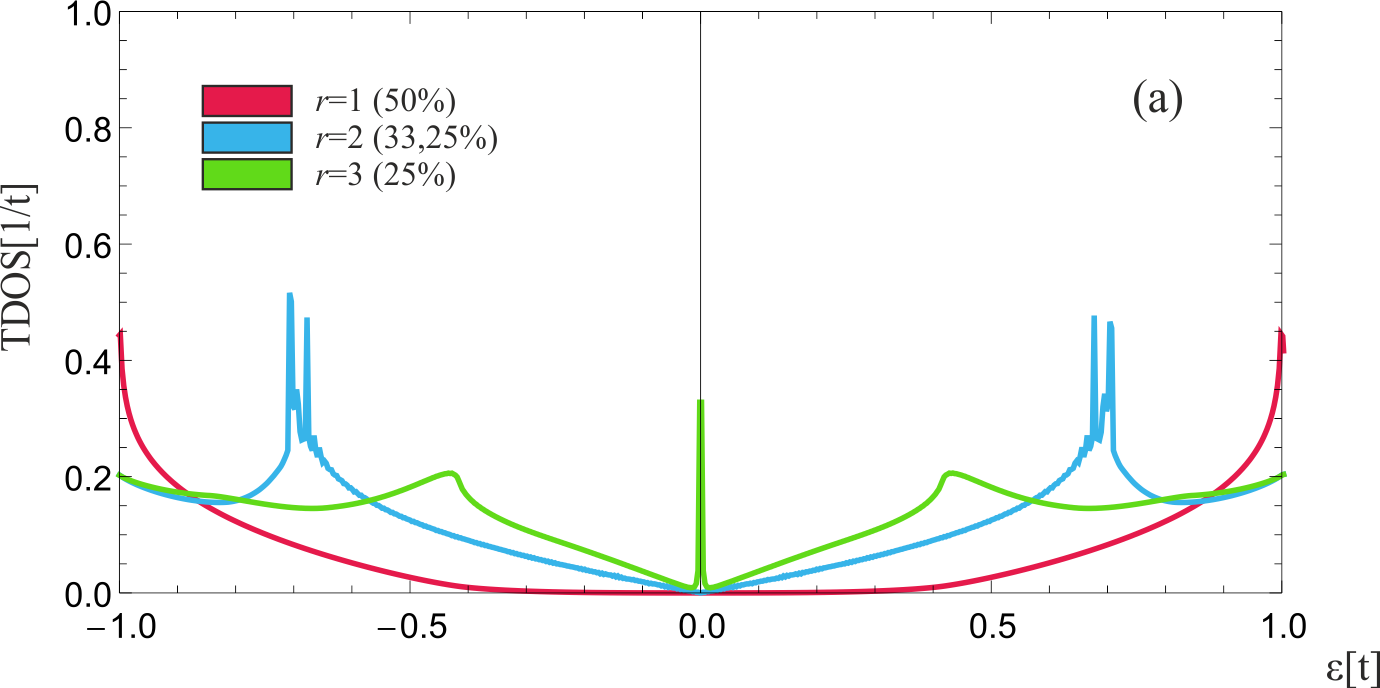}
		\includegraphics[width=8cm]{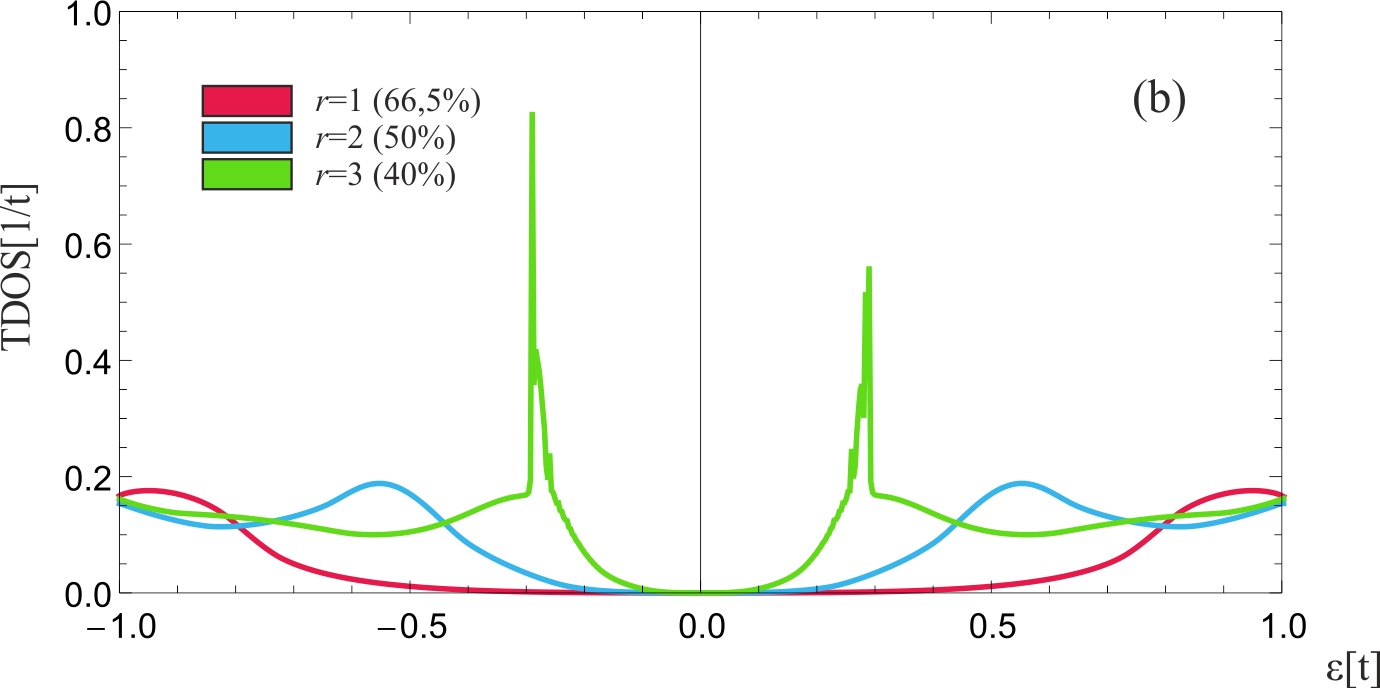}
	\end{center}	
	\caption{TDOS for a periodic location of single vacancies and divacancies with different periods. The edge state become disintegrated at 30\% vacancy concentration and higher.}
	\label{pon7}
\end{figure}

(iii) Periodic location. 
As shown above, a redistribution of the density of states between the main peak and subpeaks happens when two defects are placed at a critical distance $r\leq3$.
For this reason, we arrange single and pair vacancies with a period of one, two and three atoms between defects. The regularities originating to LDOS appear on the TDOS if vacancies are placed periodically (Fig. \ref{pon7}). 
The edge state disappears in all cases except for the location of single vacancies between three atoms, because this concentration is too low to distort TDOS (see Fig.~\ref{pon7} (a)). This result agrees with
that found in Ref.~\cite{au21} where the effect of structural defects at the graphene edges was discussed.   
Thus, the edge state becomes disintegrated when the concentration is about 30\% for the periodic arrangement of defects.

\begin{center}
	\textbf{Discussion and conclusion}
\end{center}

\begin{figure}[t!]
	\begin{center}
		\includegraphics[width=8cm]{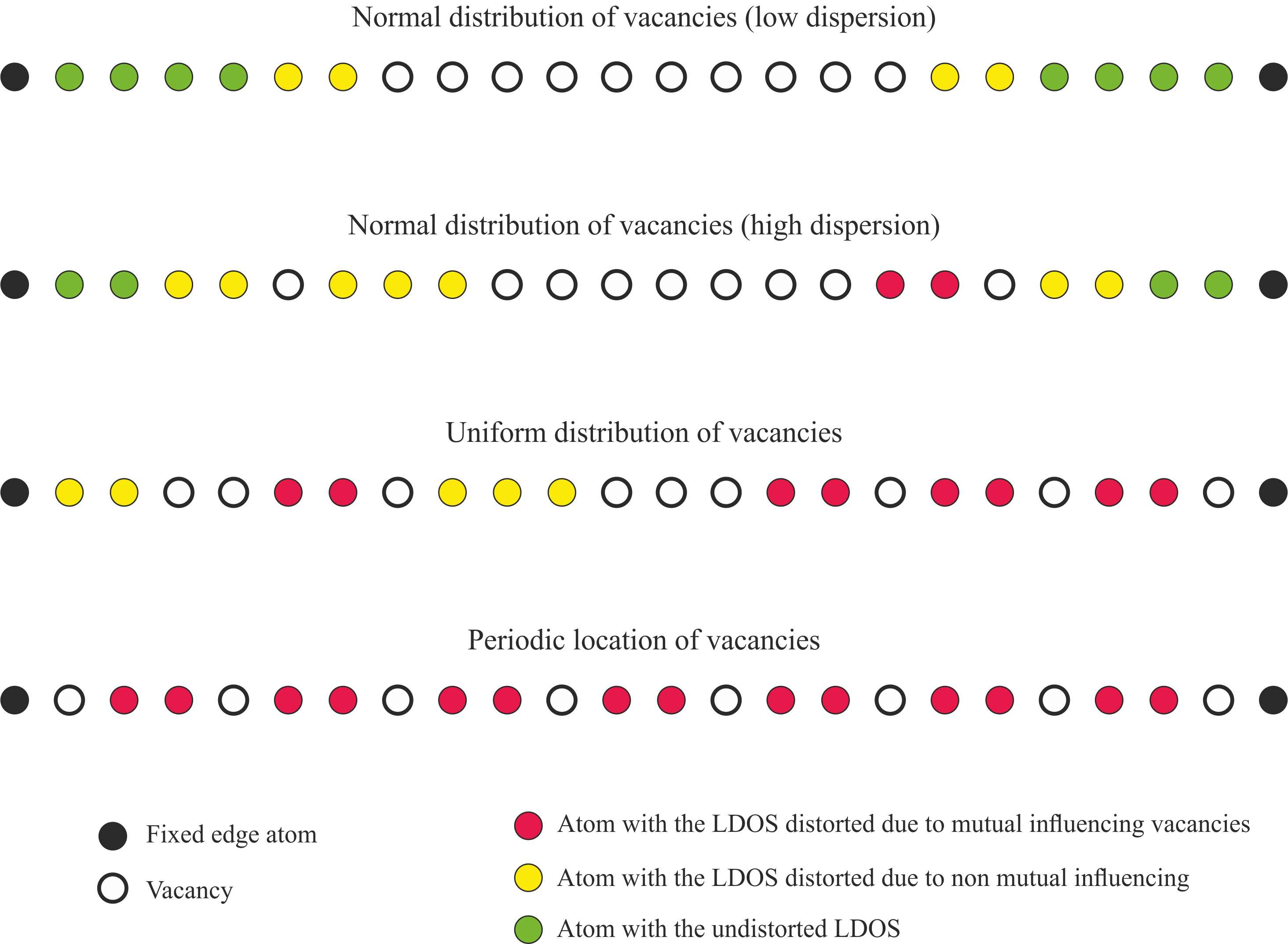} 		
	\end{center}	
	\caption{An illustrative picture of the LDOS at edge atoms for different vacancy distributions. The edge state disappears when LDOS is distorted at all atoms (for two bottom chains).}
	\label{cir}
\end{figure}

Our study shows that the stability of edge states depends critically on the position of defects. This conclusion is clearly demonstrated in Fig.~\ref{cir}.
The presence of a single vacancy, divacancy and longer arrays of defects changes the LDOS around three nearest atoms. When vacancies are located at a distance of one or two atoms, their mutual influence leads to the appearance of subpeaks in the LDOS. In this case, LDOS becomes  redistributed between subpeaks and the edge state: the peak near the Fermi level decreases while subpeaks increases. Such behavior does not occur in the case of vacancies without interference.

This means that, for example, five mutual influencing single vacancies have a stronger influence on the density of states than an array of five located in a raw defects. 	
This explains why the edge state survives when the concentration of defects is very high (70\%) in the case of a normal distribution. 
At low dispersion, the array of vacancies grows from the middle to the edges of the sheet with increasing concentration. The edge states disappear only in the case when the array reaches the edges. At high dispersion, gaps are formed in an array thus leading to 
multiple groups of mutual influencing vacancies. This explains the decrease of the central peak in TDOS.
\\
In the case of uniform distribution, there appear many mutual influencing vacancies at 50\% concentration and, as a result, the hight of subpeaks grows while the peak near the Fermi level lowers.
The edge state disappears as long as the change in the LDOS affects all atoms of the sheet.
The case of the periodic location of vacancies clearly demonstrates that namely the positions of  defects have the crucial influence on the stability of the edge state. We found that the edge state disappears for single vacancies and divacancies distributed with the period of one and two atoms. This corresponds to a 30\% vacancy concentration and higher. Notice that a similar result was obtained in Ref.~\cite{au9} for the magnetic moments which may locally vanish if two defects randomly occur to be closer than 3NN distance.
\\
To summarize, the edge state turns out to be destroyed most effectively when vacancies are located at a distance not exceeding the characteristic range of mutual influence like in the case of periodic distribution. The more vacancies influence each other, the lower concentration of defects needs to destroy the edge state. When vacancies are located according to the normal distribution, the edge state is  degraded at high concentration of defects (more than 70\%).
For the uniform distribution, the edge state is found to disappear at smaller but nevertheless quite large concentrations exceeding 50\%. The robustness of the edge state can only be ensured 
in the case of 30\% and lower vacancy concentration. However, it should be taken into account  that subpeaks appeared in TDOS may affect the productivity of molecular devices like a planar graphene-based transistor~\cite{au13}. These studies are now in progress.

\end{document}